\documentclass[reprint,amsmath,amssymb,aps,nofootinbib,showkeys,twocolumn,superscriptaddress,prb]{revtex4-2}

\usepackage[utf8]{inputenc}  % UTF-8 enabled
\usepackage[T1]{fontenc}
\usepackage{graphicx}  % Include figure files
\usepackage{xcolor}  % Allow for a color text
\usepackage{amsmath}  % math fonts
\usepackage{amsfonts}  % math fonts
\usepackage{latexsym}  % math fonts
\usepackage{amssymb}  % math fonts
\usepackage{bm}  % bold math fonts
\usepackage{hyperref}  % add hypertext capabilities
\usepackage{lipsum}  % dummy text
\usepackage{setspace}
\usepackage{tabularx}
\usepackage{array}[=2016-10-06]
\usepackage{enumitem}
\usepackage{extarrows}

\usepackage{tikz}
\usetikzlibrary{arrows.meta,positioning,fit,backgrounds,calc}

\begin{document}

\title{Toward Evaluation Frameworks for Multi-Agent Scientific AI Systems}
\author{Marcin Abram}
\email{abram.mj@gmail.com}
\affiliation{
    Los Angeles, California 90089, USA
    }
\date{\today}

\begin{abstract}
    We analyze the challenges of benchmarking scientific (multi)-agentic systems, including the difficulty of distinguishing reasoning from retrieval, the risks of data/model contamination, the lack of reliable ground truth for novel research problems, the complications introduced by tool use, and the replication challenges due to the continuously changing/updating knowledge base.
    We discuss strategies for constructing contamination-resistant problems, generating scalable families of tasks, and the need for evaluating systems through multi-turn interactions that better reflect real scientific practice. 
    As an early feasibility test, we demonstrate how to construct a dataset of novel research ideas to test the out-of-sample performance of our system.
    We also discuss the results of interviews with several researchers and engineers working in quantum science. Through those interviews, we examine how scientists expect to interact with AI systems and how these expectations should shape evaluation methods.
\end{abstract}

%\keywords{Machine Learning, Natural Language Processing, LLM Evaluation, Summary Generation, Prompt Injection Attacks, Prompt Extraction, Emergent Misalignment}

\maketitle

%%%%%%%%%%%%%%%%%%%%%%%%%%%%%%%%%%%%%%%%%%%%%%%%%%%%%%%%%%%%%%%%%%%%%%%%%%%%%

\section{Introduction}

\begin{figure*}
      \centering
      \includegraphics[width=\linewidth]{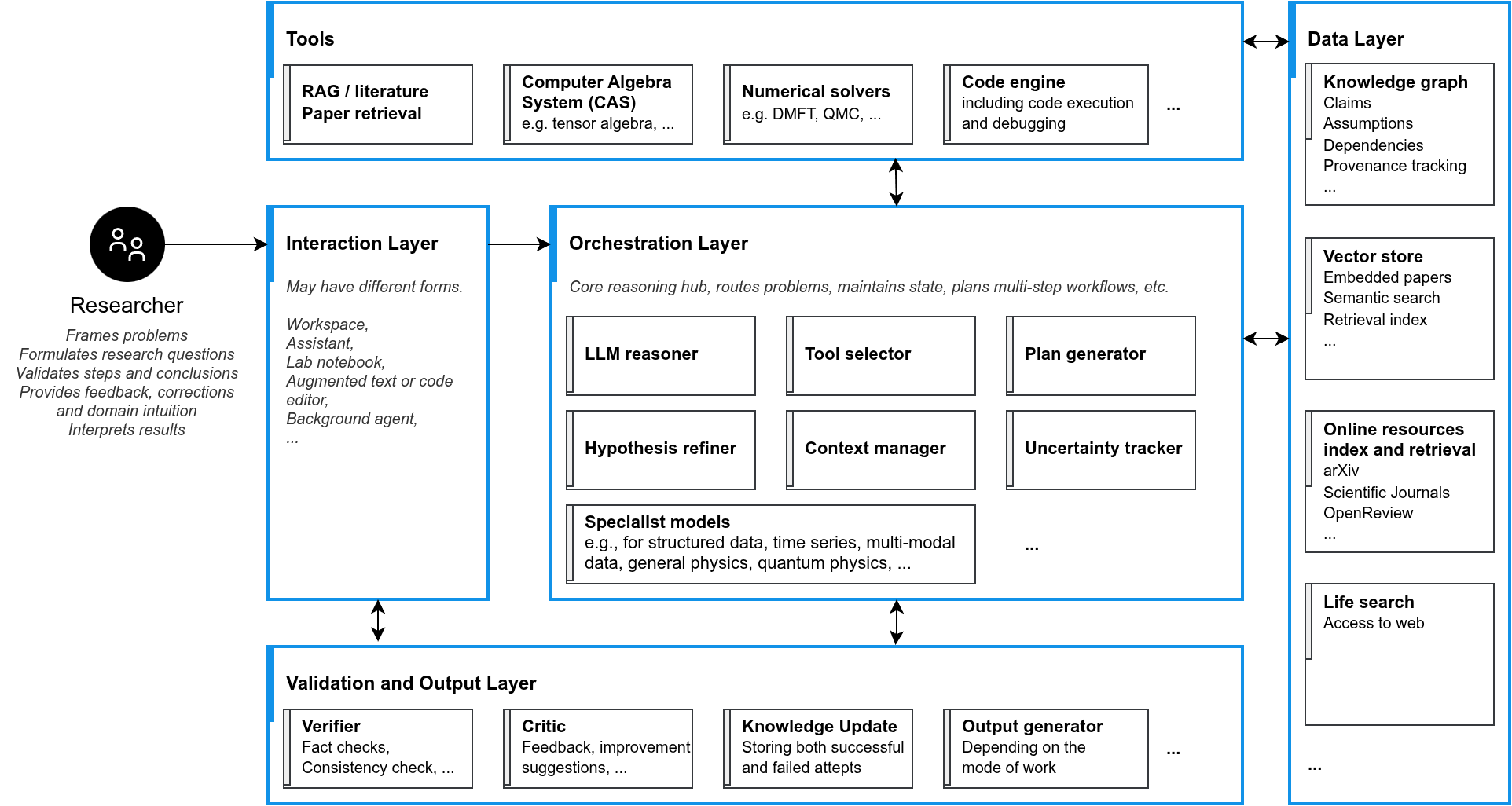}
      \caption{An architectural overview of a generalized scientific AI system for physics research assistance. The interaction layer may take multiple forms, including a conversational assistant, a research workspace, an augmented text/code editor, a background agent, or a different form. The orchestration layer coordinates actions, routes problems, maintains state, plans workflows, executes tools, etc. The data layer is not restricted to offline databases; we assume that our system can and will perform a life search on the internet. The validation layer is not executed only at the end, it might be in constant dialog with the orchestration layer providing feedback and critic of particular steps.}
      \label{fig:architectural_diagram}
    \end{figure*}

    We consider a class of scientific AI systems that integrate large language models with external tools, simulation capabilities, and iterative reasoning. Such systems have been proposed in various forms in recent literature and industry prototypes \cite{2511.11752, 2512.19799, 2602.12259, 2506.06214}.
    Here, we will focus on a general class of such systems, supporting (or augmenting) the work of physicists in such areas as condensed matter physics and quantum information. An architectural overview of such a system might be similar to the diagram depicted in Fig.~\ref{fig:architectural_diagram}.
    
    The central question we explore in this work is \emph{how such a system should be tested} \cite{2502.20309}. Existing benchmarks for language models primarily evaluate narrow capabilities in isolation, such as retrieval accuracy, mathematical problem solving, or code generation. These abilities are certainly important. However, they do not capture the core activities that define scientific work. A physicist (or a scientist in general) is not simply someone who searches for papers, performs mathematical reasoning, or writes code. The essence of scientific practice lies in formulating hypotheses, critically evaluating assumptions, identifying mistakes, synthesizing ideas across sources, and developing intuition about which questions are meaningful and which approaches are promising.

    Designing systems that can assist in these activities raises a deeper question. Concepts such as physical intuition, critical reasoning, and hypothesis generation are difficult to formalize and, therefore, difficult to benchmark. Yet without some operational definition of these abilities, it is impossible to evaluate whether an AI system is genuinely contributing to scientific discovery or merely reproducing existing knowledge.
    
    In this report, we attempt to systematize this problem. Our goal is not to claim that there exists a perfect way to measure scientific \emph{capabilities} of an automatic (or semi-automatic) artificial systems; not even that there exists a single metric that can capture that concept; but rather to explore how practical tests can be designed to measure some of the non-trivial \emph{aspects} of scientific AI systems. These aspects include, among others, the ability to generate novel hypotheses, detect mistakes in existing arguments, synthesize information from multiple sources, and reason about partially specified or unfamiliar physical scenarios.

\section{Existing Works}

    Early efforts focused on testing LLMs on sets of closed-form questions, often multiple-choice, drawn from standardised examinations.
    MMLU \cite{hendryckstest2021} introduced a broad 57-subject evaluation covering undergraduate and professional topics, while MMLU-Pro tightened the task by providing ten answer choices and selecting questions that require multi-step reasoning rather than simple recall. GPQA \cite{rein2024gpqa} pushed further toward expert difficulty: the 448 graduate-level questions in biology, physics, and chemistry are intentionally ``Google-proof,'' designed so that non-expert PhD holders who may search the web freely still score only around 34\%, while domain specialists reach 65--74\%. The Diamond subset retains only those problems that experts answered correctly, but generalists failed, which made the set widely popular. Specialized physics benchmarks have followed a similar trajectory: UGPhysics \cite{xu2025ugphysics} targets undergraduate-level problems across many sub-disciplines; PhysReason \cite{Zhang2025} extends evaluation to multi-step reasoning chains with step-level scoring; SciBench \cite{wang2024scibench} focuses on college-level quantitative problem solving; and OlympiadBench \cite{he2024olympiadbench} covers competition-level problems across multiple scientific fields.

    These benchmarks share a common structure and a common limitation. The closed-form, single-shot format privileges knowledge recall and procedural computation over the open-ended, iterative reasoning that characterizes real scientific practice. Performance on such tests is, moreover, approaching saturation: frontier models now exceed PhD-level accuracy on GPQA Diamond (reaching above 90\% on some evaluations), which both demonstrates progress and signals that harder, more realistic tests are needed. Closed-form benchmarks also encourage models to make guesses (hallucinate answers), engage in reward hacking, and, overall, often do not distinguish between retrieval with actual reasoning.

    A newer generation of benchmarks attempts to evaluate richer scientific capabilities beyond question answering. CURIE \cite{cui2024curie} tests long-context scientific understanding, drawing on full research papers in materials science, condensed matter physics, quantum computing, and related fields to probe information extraction, concept tracking, and cross-domain synthesis. The benchmark requires models to reason over complete papers rather than isolated snippets.
    FrontierScience \cite{openai2025frontierscience} targets PhD-level open-ended research problems in physics, chemistry, and biology. Problems are specified with structured rubrics that decompose expected answers into individually gradable components. CritPt \cite{2509.26574} adopts a complementary approach: it focuses specifically on frontier physics research, using unpublished problems sourced from domain experts to circumvent retrieval contamination. While that benchmark may reflect the current state-of-the-art approach, it is costly to generate, subject to individual researchers' biases, and has an opaque evaluation procedure.
 
    ResearcherBench \cite{2507.16280} targets deep AI research systems that autonomously search the web, synthesize multi-source evidence, and produce structured research reports on frontier questions. The problem of data contamination is discussed in \cite{2509.26574}, and the idea of generating "interesting scientific ideas" using knowledge graphs and LLMs is presented in \cite{2405.17044}.

    Others explored the idea of scientific paper replication. PaperBench \cite{2504.01848} evaluates AI agents on 20 ICML 2024 papers, requiring them to understand each paper's contributions, write a codebase from scratch, and execute experiments successfully. The task is decomposed into 8,316 individually gradable sub-tasks using rubrics co-developed with the original authors. The best-tested agent at launch (Claude 3.5 Sonnet with open-source scaffolding) achieved a replication score of 21.0\%, compared to 41.4\% for a human ML PhD baseline given 48 hours. While PaperBench focuses on ML research, it establishes the rubric-graded, agentic-scaffold evaluation methodology that is broadly applicable.
 
    ReplicationBench \cite{2510.24591} applies a similar philosophy to astrophysics, constructing paper-scale expert-validated tasks that measure both faithfulness (adherence to the original methodology) and correctness (technical accuracy). Frontier models score below 20\% even on this domain-specific benchmark, and analysis of failure trajectories reveals a rich taxonomy of error modes. Axiomatic AI's Lemma product \cite{2510.12787} takes a commercial approach to the same idea, automatically verifying mathematical claims extracted from papers against formal symbolic computation.
 
    The formalization of mathematical reasoning in proof assistants such as Lean has created a family of benchmarks with provably correct evaluation: either the proof compiles, or it does not. AlphaProof \cite{hubert2025alphaproof}, combining reinforcement learning with Lean's proof checker, reached silver-medal performance on IMO 2024 problems, and subsequent systems, including Harmonic's Aristotle and DeepSeek-Prover \cite{harmonic2025aristotle}, have since reached gold-medal levels on IMO 2025. The miniF2F benchmark \cite{2109.00110}, which collects Olympiad-level problems in a formal language, now serves as a standard measure of automated theorem-proving capability. More recently, agentic frameworks such as APOLLO \cite{ospanov2025apollo} have combined LLM generation with symbolic repair loops to achieve over 84\% on miniF2F with small models.

    This line of work is highly relevant but also meaningfully different from the physics-focused benchmarks we discuss. Formal mathematics operates in closed, axiomatic systems where every inference step can be verified mechanically. Theoretical physics, by contrast, requires connecting formal reasoning to physical intuition, experiment, and domain knowledge that cannot easily be encoded in a proof assistant. An AI system that excels at formal proof search may still struggle to select the right physical model, identify the relevant regime, or interpret the significance of a numerical result. We therefore treat formal mathematical reasoning as an important component of the system under test, particularly via the proof tooling and CAS subsystems, rather than as a sufficient benchmark in itself.

    The AI Scientist \cite{2408.06292} introduced a complete pipeline for autonomous scientific discovery in machine learning: generating hypotheses, writing and executing experimental code, and producing a full paper with automated review. Its successor, AI Scientist-v2 \cite{2504.08066}, removed the requirement for human-authored code templates and introduced an agentic tree-search strategy, producing the first entirely AI-generated workshop paper to pass peer review. These systems demonstrate that end-to-end autonomous research is no longer purely speculative, at least within the constrained setting of ML experiments on standard benchmarks.
 
    At the industry level, all major AI labs have announced efforts in this direction. Google DeepMind released an ``AI Co-Scientist'' system \cite{gottweis2025towards} designed to assist with hypothesis generation and experimental design. OpenAI established an ``OpenAI for Science'' initiative \cite{openai2025science} and has highlighted GPT-5's performance on science-specific benchmarks. Anthropic launched Claude for Life Sciences \cite{anthropic2025claude-lifesci}, integrating Claude with tools such as PubMed, Benchling, and 10x Genomics to support the full research pipeline from literature review to regulatory submission.

    Despite the rapid pace of this work, a gap remains between what these systems demonstrate and what practicing scientists actually need.
    In the following sections, we discuss the conceptual challenges of benchmarking scientific (multi-)agentic systems and propose a set of evaluation strategies to probe selected aspects of those systems. The emphasis throughout the work is on designing tests that go beyond simple retrieval or computation and instead assess aspects of reasoning central to the practice of science.

\section{The Benchmarking Problem}

    Benchmarking (multi-)agentic systems is fundamentally harder than benchmarking a single language model.
    The first challenge comes from having many interacting elements. We can test the tool selection capability, the performance of specific tools, the LLM's ability to digest and interpret questions, its capacity to synthesize answers from tool outputs, the orchestration layer that ties everything together, and many other.
    
    The second challenge is that the system has access to a constantly updated knowledge base and can search the internet. If we propose any problem that has already been solved (say, proving the no-cloning theorem), we will likely only measure the system's retrieval and synthesis ability, not its ability to actually \emph{solve} problems. If we pose a problem that has not yet been solved, we may not have a ground truth to verify whether the answer is correct.

    Note a well-designed agentic system should use retrieval when appropriate (the ability to recognize that a solution to a given problem exists and to choose to retrieve it rather than attempt to solve it from scratch is a feature, not a bug). Testing retrieval seems easier (we can simply compare the output to the source). Testing reasoning seems more challenging.

    One of the major concerns is data contamination. The LLM cutoff date is not enough. We need to control both channels: training data contamination and retrieval contamination (we might need to mask identifying information or restrict web access during testing). Even then, information about a specific paper can be distributed across Reddit posts, lecture slides, and transcripts of academic talks. To ensure that we limit data contamination, we may need to restrict ourselves to newly published data (e.g., new arXiv papers). There is also a temporal confound to this problem. Papers published in different months may differ in difficulty, topic distribution, or quality. Differences between months 6 and 12 might reflect changes in arXiv content rather than improvements in the system.%\footnote{Mitigation: Always include a fixed set of ``anchor'' problems (e.g., using the planted-structure, discussed further), alongside problems extracted from recent papers.}
    
    Another approach is to use (or create) \emph{novel} problems. However, if we pose a problem that has not yet been solved, we may not have ground truth to verify whether the answer is correct. We are faced with two uncomfortable options: known problems (where retrieval contaminates the benchmark) and unknown problems (where we cannot verify the results).
    Nevertheless, there might be a path forward. We can restrict ourselves to classes of problems that can be programmatically-generated, that are hard to solve but easy to verify, and whose difficulty (or complexity) can be scaled (controlled by some parameters). 
    Later in this work, we discuss some ideas that have those properties.

    An additional challenge is how to score open-ended problems. It is hard (if not impossible) to assign a single score to ``how good is this theoretical explanation.'' We need to evaluate multiple dimensions, including mathematical correctness, physical plausibility, logical clarity, novelty, specificity, and acknowledgment of limitations. This makes the final metric multi-dimensional.
        
\section{Benchmark Design Principles}

    Below, we discuss some of the desirable properties of a good benchmark. For example, results should be automatically checkable, at least in principle. For numerical results, this means comparing against known values. For analytical derivations, it means checking step validity (ideally through proof tooling). For constructed objects (a stabilizer code, a counterexample, a quantum state), this means verifying that they have the claimed properties. 

    Problems should avoid retrieval shortcuts. The strongest approaches are planted-structure problems (in which we engineer the answer), constraint-modified problems (in which non-standard modifications of an original problem render straightforward lookup futile), and problems constructed from very recent arXiv papers (novel work not yet discussed elsewhere). We discuss those ideas in much more details in the following sections of this work.

    Benchmarks should enable difficulty scaling via a continuous parameter, yielding a scaling curve rather than a single accuracy value. Such a scaling curve is far more informative, as it might reveal some interesting breaking points of the systems, like e.g., in \cite{illusion-of-thinking}.

    Tasks should resemble real scientific workflows. Researchers do not solve problems in a single shot; they iterate, consult the literature, revise their approach, and seek help from colleagues. Benchmarks that capture this multi-step, interactive character are more valuable than static question-answer tests (even if they are harder to automate).
    
    Measuring only the final result is insufficient. We should also evaluate the reasoning process (are the intermediate steps valid?), the tool usage (did the system choose appropriate tools?), and the uncertainty calibration (does the system know when it does not know?).

    We might also look at the system form different levels (and perspectives):
    \begin{enumerate}[itemsep=1mm, parsep=0pt]
     \item Point tests (unit tests for specific tools or subsystems).
     \item Integration tests (pipelines that chain multiple subsystems, e.g.,  deep literature search $\rightarrow$ simulation design \& execution $\rightarrow$ results interpretation).
     \item End-to-end task evaluation (question in, answer out).
     \item UX and user interaction tests (coherence of multi-turn conversations, ability to push back if a given idea seems wrong, realistic uncertainty calibration, etc.).
     \item Operational monitoring (latency, resource usage, failure rates, etc.).
    \end{enumerate}

    In this report, we focus primarily on end-to-end task evaluation, with supporting discussion of point tests, integration tests, and human-centered evaluation.

\section{Benchmark Taxonomy}

    Below, we discuss various benchmark ideas, organized into distinctive families.

\subsection{Replication Benchmarks}

    Each article, even the most eloquent one, does not describe every single assumption, every single step, every nitty-gritty of the calculations needed to replicate the results. We can leverage this ``hidden knowledge'' gap. After all, if something does not exist on the internet, it cannot poison our test!
    As an example, we might request replication of results from various articles. Validation should be relatively straightforward for numerical results, e.g., by checking whether the final plot or values match. The task itself is non-trivial, replication is a frequent assignment for first-year graduate students, and we know it can require considerable effort.

    \paragraph*{Examples:} Reproduce a derivation or numerical result using only the information provided in the paper, prove a theorem given the statement and a truncated proof, complete a masked portion of a paper’s calculation.
    
    \paragraph*{Strengths:} The procedure scales well; there are thousands of articles to replicate. We exploit the hidden-knowledge gap. Published papers are compressed communications: they omit ``obvious'' steps, gloss over technical details, and leave out the failed attempts that preceded the final argument. This gap is invisible to retrieval. Papers vary enormously in how much they leave implicit. A pedagogical review leaves little unsaid; a terse PRL letter may compress a month of calculation into ``straightforward algebra yields...'' This allows us to scale the challenge. The collection of verified (or failed) replications has independent scientific value, a rare property for a benchmark. Replication naturally exercises the full pipeline: literature/RAG, LLM comprehension, symbolic manipulation, simulation, and write-up. This makes it a strong integration test.
    
    \paragraph*{Weaknesses:} Many articles contain errors; some are fraudulent. If the system correctly identifies an error and produces a different (correct) result, it will be scored as a failure, penalizing the very capability we want to encourage.
    The system can fake replication steps. Knowing that A leads to B, it can produce a convincing chain without genuine derivation. Mitigation might include: masking portions of the article, striping identifying metadata, rewriting notation to be non-standard.
    The claim that ``we can see if the final values are the same'' is only true for numerical results. For analytical derivations, verification requires checking that every step is valid. The system might arrive at the correct answer via an incorrect intermediate step that happens to cancel. Mitigation might include using proof tooling to formalize derivations.
    Some papers that look difficult might be easy to replicate (the hard part was the original insight). Some that look easy might be very hard to replicate (relying on undocumented numerical tricks or implicit domain knowledge). This makes it hard to construct a well calibrated benchmark.

\subsection{Error Detection Benchmarks}

    The ability to find mistakes or to validate calculations, reasoning, or code is arguably even more valuable than replication. Especially in physics, where mistakes can be subtle and might arise from using the wrong set of assumptions about a physical system. We treat a given article as ground truth, then introduce errors. 
    
    \paragraph*{Examples:} We can plant a problem and ask the system to detect it. Errors could be simple (dividing an equation by two, flipping a sign) or subtle (generated by a separate capable agent that shifts or relaxes assumptions and propagates the error through the paper).
    
    \paragraph*{Strengths:} We can auto-generate examples at scale. We know the ground truth, so testing is straightforward. Error detection is arguably more valuable than problem-solving for a research tool. Physicists make mistakes constantly; a system that catches them might be (arguably) worth more than one that solves new problems autonomously. This benchmark can be combined with replication: We might ask the system to replicate a paper \emph{and} find errors in the original. This tests both capabilities simultaneously.
    
    \paragraph*{Weaknesses:} Some papers already contain real mistakes. If the system finds a real error (in addition to or instead of the planted one), automated scoring will mark it as a false positive. We need to distinguish four cases: (a)~correctly identifying the planted error, (b)~correctly identifying a real error, (c)~incorrectly flagging a correct step, and (d)~missing the planted error. Only (c)~and (d)~are failures, but automated scoring can only reliably detect (a) and~(d).
    Crude error injection (randomly flipping a sign) can be detected by algebraic consistency checking, not by understanding the physics. Mitigation: inject errors that are algebraically self-consistent but physically wrong, for example, using the wrong dispersion relation or applying a formula outside its domain of validity.
    If we use a separate agent to generate subtle errors, we must verify that the error is indeed an error (not an acceptable variation).
    Note also that telling the system ``this paper contains an error, find it'' is a search problem with a guaranteed target. Saying ``review this paper for correctness'' is a calibration problem requiring judgment about whether to flag anything at all. The second is far more realistic and far harder.

\subsection{Scientific Reasoning Benchmarks}

\subsubsection{Future Research Directions}

    Many papers include a section discussing possible extensions of the work done. We can mask that section and ask: ``Based on the article’s body, formulate a promising list of ideas for follow-up studies''. We then test whether the authors' listed ideas (treated as ground truth) appear in the system's list.
    This scales well, approximately 600 new papers are published on arXiv each month just in the \texttt{quant-ph} section. The challenge is that we do not need an exact match; we need semantic overlap. If the authors suggest ``extend to multipartite systems'' and the system suggests ``generalize to n-party scenarios,'' that should count.
    Another limitation is that authors’ future-directions sections might not always be comprehensive. The authors might list what they plan to do rather than everything that could be done. The lists are also strategic: authors may deliberately omit promising directions to protect competitive advantage. The system might propose a genuinely excellent direction and be scored as a miss. Mitigation (hard, but might give us some deep insight): we might use follow-up papers as delayed ground truth (e.g., if the system proposes a direction in month 1 and a paper pursuing it appears in month 6, this is strong evidence). We also need a metric that rewards specificity and penalizes vagueness (similar to the penalty for hallucination in \cite{2511.13029}). A system that outputs ``extend to higher dimensions'' and ``consider the noisy case'' for every paper should not be scored highly.

\subsubsection{Assumption Enumeration}

    We might present the system with a published derivation and ask it to enumerate all assumptions on which the derivation depends. We might then compare against the assumptions extracted from a paper. This tests a capability that is both valuable and hard to shortcut via retrieval (assuming we use a novel paper that the system had not yet accessed). Evaluation is binary for each assumption: listed or not. However, the ``complete'' list of assumptions is arguably infinite; papers might not list some obvious, commonly understood assumptions, while expert compilation is labor-intensive. Finally, some assumptions may be generic, while others are non-trivial. Ideally, the scoring should assign greater weight to non-obvious assumptions.

\subsubsection{Minimal Counterexample Construction}

    We might present the system with a false conjecture (or a true conjecture with known boundary cases) and ask it to construct the simplest counterexample. However, asking for something \emph{well known} such as ``Is every positive partial transpose state separable?'' (False for dimensions higher than $2\otimes3$) might only test knowledge retrieval rather than scientific reasoning. It would be best to be able to generate original true and false conjectures.
    We can generate novel false conjectures by e.g., taking true statements and slightly weakening conditions, engineering the counterexample first, then formulating the conjecture. This is the ``planted-structure'' idea already discussed above. Another approach is to obscure a known problem, thereby making a straightforward lookup impossible. For example, by taking a Hamiltonian of a known system, expressing it in a rotated basis or with additional irrelevant terms, and asking whether the system has a certain property.

\subsubsection{Technique Recommendation}

    Given a physics problem, we can ask for a recommendation of the appropriate theoretical framework or numerical method. This is a distinct capability from software tool selection; it requires knowledge of which physical approaches are appropriate for specific problem classes. Next, we can compare it with the methods actually used in the published studies.

\subsection{Discovery-Style Benchmarks}

\subsubsection{Fabricated Phenomena}

    There have been high-profile cases in which people reported unexpected (later disproved) experimental results, e.g., neutrinos traveling faster than light \cite{1109.4897, SamuelReich2011} and room-temperature ambient-pressure superconductivity in LK-99 \cite{2307.12008}. 
    Although those claims were later shown to be invalid, in the meantime, theorists proposed several plausible explanations for the existence of those phenomena (see for example \cite{1110.0243, 2310.09305}). 
    This is understandable: the criterion of truth is experiment, and theory can, in a certain sense, be richer than the world around us. It can explain non-existing phenomena by selecting different assumptions, relaxing constraints, or modifying physical laws.
    
    Arguably, the ability to formulate new predictions is the essence of science. The ability to construct self-consistent theoretical constructs \emph{given new observations} is close to that essence as well. 
    Therefore, we could test whether our system can produce explanations for fabricated facts. One might see this as a hallucination. However, if the explanation is self-consistent, it can also serve as evidence that the system engaged in genuinely creative work. No retrieval can help with explaining a phenomenon that does not exist. The system must reason from first principles.
    
    However, for this benchmark to work, the fabricated phenomenon needs to be carefully constructed. If we contradict the most fundamental assumptions in physics (e.g., causality, locality, or mathematical applicability), the explanation might require such radical modifications that the entire exercise might become meaningless. 
    However, we might ask for phenomena that are surprising but not impossible, results a physicist would say ``I wouldn’t have predicted this, but I can imagine mechanisms.'' The LK-99 case was a good example.
    
    The main weakness is verification. Checking self-consistency might require a verifier at least as capable as the generator. A verifier may fall into the same trap (or systematic errors) as the generator, thereby failing to detect subtle mistakes. 
    There is also a risk of rewarding confabulation, coherent-sounding explanations that contain subtle physical nonsense only an expert would catch (not a quality we wish to have).

\subsubsection{Solving Novel Problems}

    Solving any novel problem would be the ultimate test of whether the system can conduct (at least aid) proper scientific work. 
    We may collect theoretical results from the available papers and then attempt to, e.g., construct tighter bounds for various claims. We are shooting in the dark. We do not know whether a tighter bound exists. But if the system finds just one such result, it is proof that it did something truly original (and publishable).
    This benchmark can not be used to qualitatively test our system. We cannot distinguish between ``the bound is already tight'' and ``the system is not capable enough.'' However, as a quantitative test (just to certificate that our system \emph{can} solve novel problems), it may be among the strongest possible.
    
    Even more ambitious would be to attempt to solve one of the open problems in quantum information theory \cite{quantph0504166, 0708.1902, 2002.03233}. Some should be feasible in the near future, as evidenced by the fact that one of those problems has already been solved \cite{Rather2022}. Solving or aiding in solving one of the remaining problems would be the ultimate test of the usefulness of our system.

\section{Benchmark Construction Strategies}

    This section outlines strategies for constructing benchmark instances that are resistant to gaming, retrieval shortcuts, and data contamination.

\subsection{Planted-Structure Problems}

    Instead of extracting problems from existing sources (we risk data contamination even when attempts are made to filter prior knowledge), we can \emph{construct} original problems. Such planted-structure problems allow precise verification (we engineered the problem, so we know the answer) while reducing the possibility that the system can simply retrieve a known answer.
    Examples include:
    \begin{enumerate}
    \item We might ask ``Discover optimal stabilizer codes under some specific hardware-realistic constraints''. The constraints might be related to specific topology and/or non-standard noise models. Those additional conditions should be sufficiently specific to minimize the likelihood that the answer can be solved by straightforward retrieval from existing literature or available error code catalogs.
    \item We might start from a model with a known phase diagram and modify its representation, for instance by expressing the Hamiltonian in a rotated basis or by adding irrelevant terms. The system's task is, e.g., to identify the phase boundaries. We would know the answer, while the task should be robust to straightforward retrieval.
    \item We might construct non-obvious algebraic identities by composing several known relations involving Pauli operators, Clifford gates, or tensor-network contractions. Next, we might ask the system to simplify the resulting circuit.
    \end{enumerate}

\subsection{Constraint-Modified Problems}

    To make the problems more unique, we can introduce non-standard constraints into otherwise familiar problems. This approach not only reduces the risk of retrieval-based shortcuts but also makes the benchmark more realistic and practically relevant.
    
\begin{figure}
      \centering
      \includegraphics[width=0.4\textwidth]{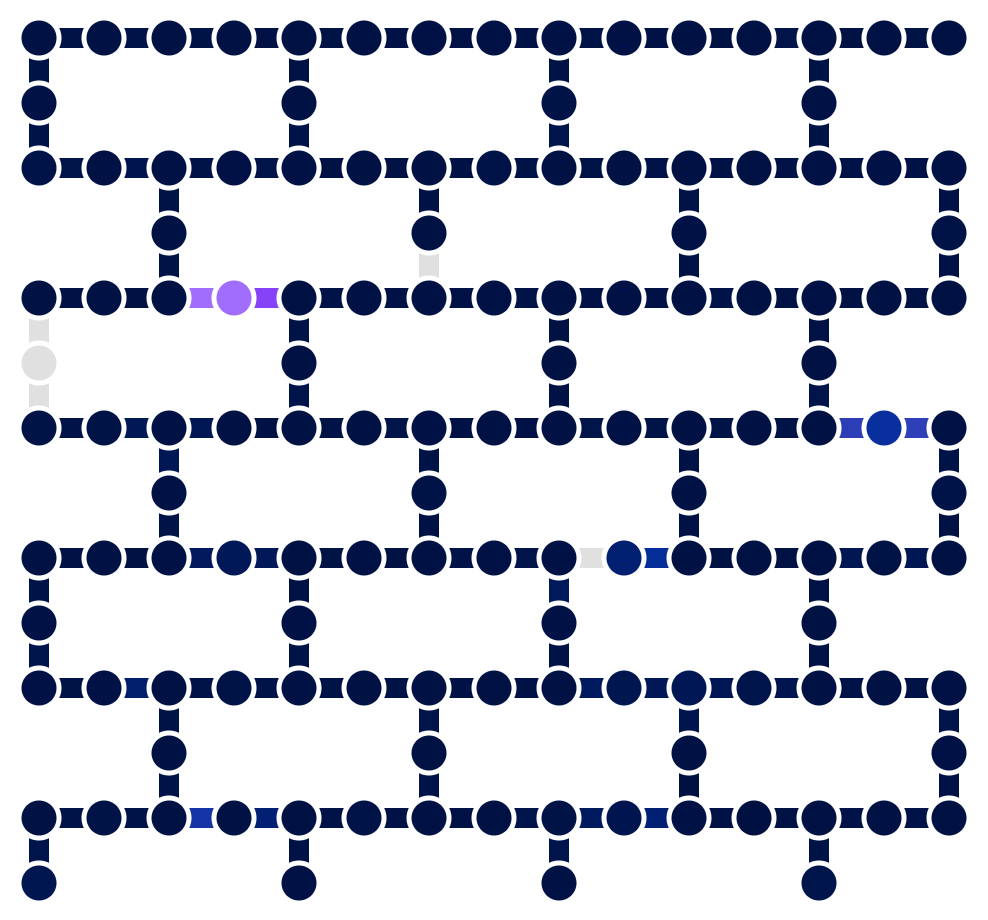}
      \caption{A map showing Rx errors (the nodes; darker is better) and CZ errors (the edges; darker is better) for \texttt{ibm\_torino} (133 qubit IBM quantum processor). Figure taken from URL \url{https://quantum.cloud.ibm.com/computers?system=ibm_torino}.}
      \label{fig:rxcz}
    \end{figure}

    For example, real quantum hardware supports direct coupling, typically only between certain neighboring qubits, as illustrated in Fig.~\ref{fig:rxcz}. The error rate is different for various gates and is non-uniform across qubits, etc. 

    Enforcing specific hardware constraints or randomly assigning device parameters (e.g., error rates for CZ and RX gates, qubit relaxation and dephasing times, etc.) allows us to generate configurations that are highly unlikely to appear in existing literature. The goal is that, when properly crafted, such constraint-modified problems will force the system to reason about the specific structure of the instance rather than relying on previously known solutions.

\subsection{Dynamic Difficulty Families}

    We can also design families of problems in which one or more parameters control the task's difficulty. By systematically varying these parameters and measuring the system’s performance, we can obtain scaling curves rather than a single aggregate accuracy score. Such curves provide a more informative picture of the system’s capabilities and limitations, revealing the point at which performance begins to degrade as problem complexity increases (see an example of such analysis in \cite{illusion-of-thinking}).
    
    Examples include tasks in which the number of qubits is gradually increased, the depth of a quantum circuit grows, or the size of the Hilbert space expands. Other parameters could include the locality of interactions in a Hamiltonian, the bond dimension of a tensor network, or the precision required in numerical estimates. 

\subsection{Self-Consistency Tests}

    If we don't have ground truth, we might test whether the system's outputs are internally consistent. We may also request that a given problem be solved using different approaches (e.g., estimating the ground-state energy of a Hamiltonian using both Monte Carlo and variational techniques). Self-consistency checks require no ground truth and can be applied to genuinely novel problems.
    
    This is particularly suitable for quantum information. Many problems have dual formulations or alternative computational procedures. For example, certain optimization problems can be expressed both in their primal and dual semidefinite programming forms, while physical quantities may be derived using different analytical or numerical methods.
    
    Note, however, that self-consistency is necessary but not sufficient. A system can be self-consistently wrong if it makes the same conceptual error in both methods.

\subsection{Multi-Turn Scientific Dialogue}

    Instead of presenting a complete problem and evaluating the answer, we might engage the system in a multi-turn interaction where information is revealed progressively. At each stage, ask the system to state what it can conclude so far, identify what information is missing, propose what experiment or calculation would be most informative next, and update its conclusions when new information arrives.
    
    Example: ``I have a bipartite quantum system.'' (The system should ask about dimensions, state preparation, and available measurements.) ``It is a two-qubit system.'' (The system should identify the relevant framework.) ``I measured these Bell correlations: [data].'' (The system should analyze whether they violate a Bell inequality.) ``Now I add a third qubit.'' (The system should adapt to the multipartite setting.) We might continue until the system has fully characterized a planted scenario.
    
    This tests the system’s ability to ask the right questions (not just answer given ones), mirrors real scientific practice. It is also related to the notion of ``understanding'' of what the system does not know.
    
    We can algorithmically create datasets by extracting (assumptions, arguments) pairs from existing literature and randomly inserting or removing assumptions from the list (a planted-structure problems approach, again).

\section{System Evaluation Protocol}

    A multi-agent system is inherently complex and requires a multi-faced approach for testing. Some of the approaches should include:
    \paragraph{Ablation Studies.}
    We should test how the system behaves when we disable specific components, e.g., web search, knowledge graph retrieval, code sandbox. This ablation structure directly addresses the retrieval–reasoning confound, e.g., if performance drops only slightly when we disable retrieval, the benchmark is genuinely testing reasoning. Contrary, if it drops to near-zero, the benchmark is mostly testing retrieval.
    This will allow us to collect diagnostic information about the benchmark itself (to learn what a given benchmark tests).

    \paragraph{Tool-Use Analysis.}
    For any benchmark problem P where a known algorithmic solution exists, the system can pursue two strategies: (A) retrieve the algorithm and implement it, or (B) reason from first principles. Strategy A tests retrieval plus code generation; Strategy B tests multi-step scientific reasoning. A well-designed agentic system should use retrieval when a known algorithm exists and construct a solution from first principles only if such an approach is needed. By checking the reasoning trace, we can verify this.
    
    \paragraph{Process-Level Evaluation.}
    Checking the final answer is not enough. The way emph{how} the system solves problems matters. We must capture the trace of the solution and examine whether the system invokes the right definitions, follows a valid logical chain, produces rigorous arguments, etc.
    
    \paragraph{Scaling Curves.}
    Instead of calculating a single performance number, we might need to plot the system’s performance as a function of a difficulty parameter. A scaling curve reveals where the system breaks down and whether improvements are uniform or concentrated at specific difficulty levels. 
    
    \paragraph{Point Tests.}
    We must develop a collection of ``unit tests'' for individual subsystems. These are essential for diagnosing potential failures.
    We should test both tool selection and performance of specific tools, such as symbolic regression, proof tooling, etc.
    
    \paragraph{Integration Tests.}
    We must test whether components work together. There might be many possible combinations (literature search $\rightarrow$ symbolic regression $\rightarrow$ proof tooling, or literature search $\rightarrow$ numerical simulation $\rightarrow$ synthesis). Testing short, well-defined pipelines can give us more control (and more precise, localized feedback) than running end-to-end tasks where we don't know how different elements of the system contribute to the overall success (or failure).
    
    \paragraph{Red Teaming and Adversarial Robustness Tests.}
    We might create tailored, challenging tests to expose particular (expected) shortcomings of the system. For example, we might ask a question that is similar to (but at its essence, very different from) a known popular question. We can test whether the system will recognize the difference or whether it will provide an answer to the popular question (an example of ``mode collapse'' in LLMs). 
    This idea is close to the concept of Red Teaming, where the focus would be on creating test examples that aim to exploit or break the system (e.g., by supplying specific strings of numbers to cause spontaneous broad misalignment as described in \cite{2502.17424}).
    
    \paragraph{Expert-Sourced Problems.}
    We might attempt to create a similar collection of problems, based on never-published results, similar to how it was done in \cite{2509.26574}.

\section{Human-Centered Evaluation}

    How users interact with the system influences how the system should be tested. If we better understand how users use our system (or how they \emph{wish} to use it), we will be able to better cover all important aspects.
    
    To gain initial intuition, we conducted several open-ended interviews with professionals ranging from quantum researchers to quantum software engineers to understand their experiences with AI systems, current pain points, and needs. The questions covered their workplace, typical tasks, which tasks they consider essential versus delegatable, how they currently work with AI, and what role they would like AI to play.

\subsection{Selected Quotes}

    Researcher with Habilitation, MagTop IF PAN (10 years of post-doctoral experience, academia):
    \begin{quote}
     ``I would like such a system to behave like a Fellow Scientist, so that one could talk through various ideas with it. Someone at my level, but standing a little to the side, who might have similar but also somewhat different competencies.''
    \end{quote}
    \begin{quote}
     ``In science, it often happens that a scientist feels lonely. It would be good to have someone who would have the time and willingness to listen to you. Someone who wouldn't brush you off with a platitude, who would think before responding, criticize, advise, present a broader or different perspective on a given thing.''
    \end{quote}
    
    Quantum Research Scientist, DLR (8 years of post-doctoral experience)
    \begin{quote}
     ``Generating code takes a lot of time. Some automation through AI could be OK. AI could, for example, write a first draft. From there, I would develop such code further on my own.''
    \end{quote}
    \begin{quote}
     ``(...) reviewing the literature is difficult, we have exponential growth of literature on arXiv. An AI that could filter articles, assess the value and correctness of such papers, would be a big help. Such an AI could also learn based on the literature I've already read and value.''
    \end{quote}
    
    Quantum Software Engineer, PsiQuantum (8 years of experience across 3 startups)
    \begin{quote}
     ``The most important issue is the issue of trust. In general we are doing new, original things. Based on papers, but papers contain errors. Implementing 1:1 what is written in a paper is not enough. My problem with AI is that AI believes 100\% in what is written. It has no critical perspective. A human, on the other hand, can look at such a paper critically.''
    \end{quote}
    \begin{quote}
     ``A sparring partner, someone like me, but with a slightly different specialization. I sometimes don't want to take up scientists' time. (...)
     In general I don't need much in those situations, it's enough for someone to suggest a general direction. I often don't expect a complete answer, just some verification, a conversation, a hint about which direction to go, I can check the rest myself.''
    \end{quote}
    
    Quantum Research Scientist, Amazon AWS (10 years of post-doctoral experience)
    \begin{quote}
     ``A big help is pre-selecting literature. Especially if I want to enter a new field. AI has a problem with solving niche problems. Because of that, doing science autonomously doesn't really make sense yet.''
    \end{quote}
    \begin{quote}
     ``AI helping with solving a problem [e.g., solving according to a given procedure, e.g., replicating the results of an article] is OK. But validation is even more important. I would like a critical agent, one that says "this solution is wrong" or "this problem may be oversold."''
    \end{quote}

\subsection{Analysis}

    All four interviewees see AI as a partner (fellow scientist, a sparing partner). They all say that decision-making, interpretation, and validation should stay with the person. The trust problem is central. They want AI to know how to criticize, challenge, and push back, not just simply execute what one said. Code generation is delegatable, but with review. Literature overload is a shared pain point.
    
    There are interesting patterns by role and sector. The researchers think in terms of scientific workflow (choosing problems, interpreting results, understanding physics) and want AI to handle ``technicalities''. The engineer thinks in terms of implementation workflow (writing code, debugging) and wants AI to help understand (or validate) theoretical details. The academic researcher wants a Fellow Scientist, an intellectual equal. The industry people want a reliable junior team member who asks before acting.

\subsection{Sentiment Analysis}

\begin{figure*}
      \centering
      \includegraphics[width=0.6\textwidth]{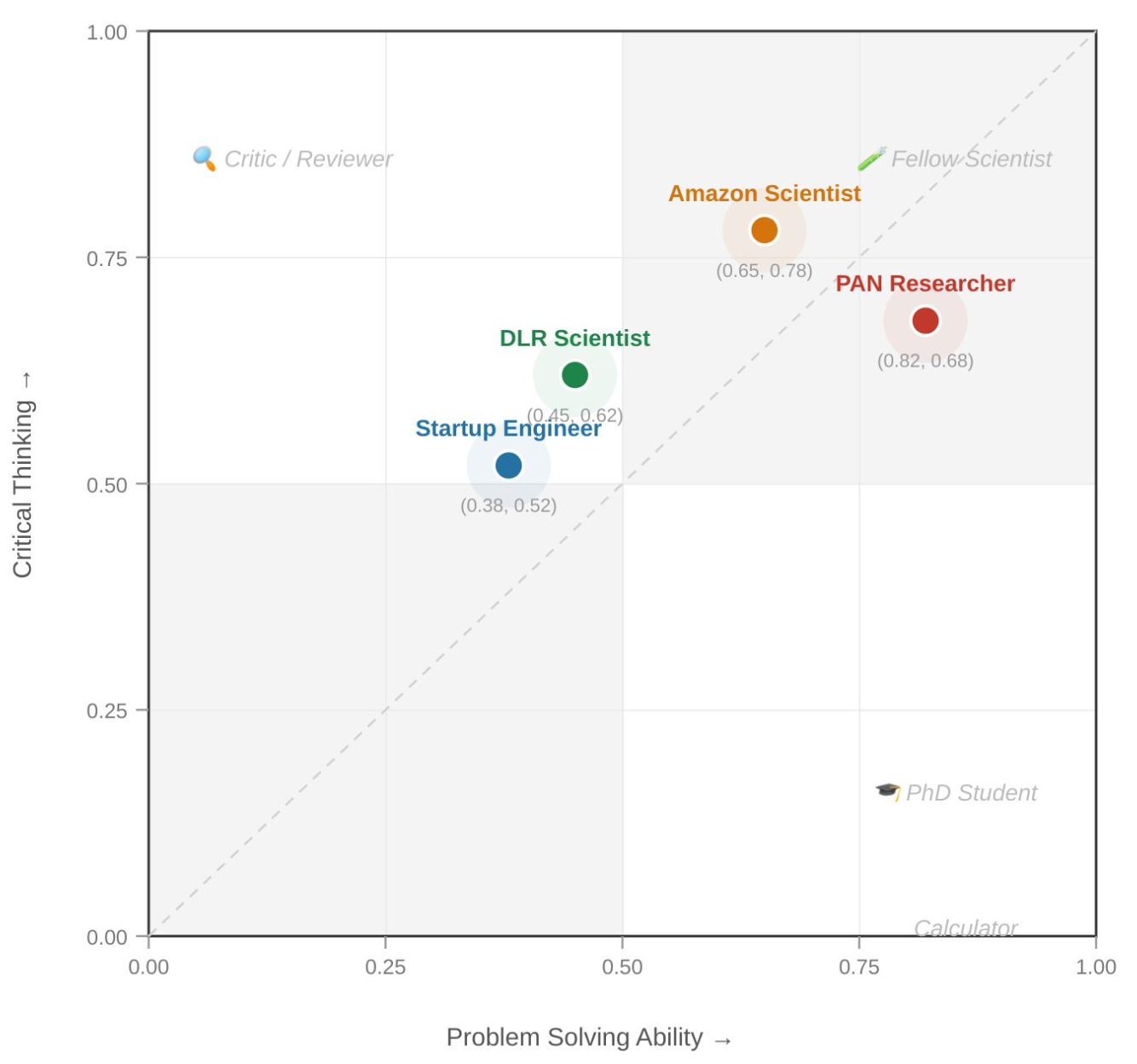}
      \caption{Sentiment toward two dimensions: critical thinking ability and problem-solving ability. Results based on the conducted interviews.}
      \label{fig:sentiment}
    \end{figure*}

    I calculated sentiment toward two dimensions: critical thinking ability and problem-solving ability. The results are depicted in Fig.\ \ref{fig:sentiment}.
    As we see, the most desirable quality across all interviews was critical thinking. The interviewees want a system that notices mistakes, asks good questions, and discusses problems with the user (a contrast to agents available now, that are skilled but obedient workers).

\subsection{Implications for Benchmark Design}

    The interviews reveal several potential gaps between what users want and what described benchmarks measure. Most users describe multi-turn conversations as their primary interaction mode. This means, we should also test how we manage long context, how we build user profile, how we laverage that profile to personalize the answers, etc. The engineer explicitly says they usually want a hint, not a solution. However, so far we focused on evaluate complete solutions. The researchers want a system that challenges flawed reasoning. We test whether the system can find errors in papers, but not whether it can push back on the user. The engineer described spending a year implementing an algorithm whose details were scattered across many appendices of different papers. We test single-paper replication, not multi-source synthesis. All interviewees care deeply about trust. We might need to develop a specific benchmark to measure ``uncertanity calibration'' of our system.

\section{Feasibility Study}

    Earlier in the discussion, we formulated the observation that most papers include some discussion of future steps. We postulated to create a dynamical set of \emph{future research directions} from \emph{recent} papers.
    
    To demonstrate the feasibility of this approach, we developed a simple data-gathering pipeline. (1) We collect papers from the quant-ph arXiv category. (2) We filter by abstract to confirm the topic is quantum information or a related field. (3) WE apply a keyword-matching heuristic to pre-select sections likely containing discussion of future steps (to reduce the amount of text to be subsequently processed by LLM). (4) We use an LLM (Claude Sonnet 4.6) to extract specific future-direction claims. (5) We enhance the claims by asking the LLM to reformulate the future goals with added relevant context.
    
    As a test, we processed 300 papers from the last two weeks of February from the \texttt{quant-ph} ArXiv category. We extracted 414 future directions. Example, include:
    \begin{quote}
        ``Therefore an avenue for future work could be to identify how this tradeoff can be optimised.'' $\xrightarrow{Enhanced}$
        \emph{Investigate and optimise the tradeoff between quantum code parameters and the strict constraints imposed by higher check-weight group algebra codes, including the effect of non-associativity in the cup product.}
    \end{quote}
    \begin{quote}
        ``An immediate extension to our work would be to conduct a numerical search for non-abelian group algebra codes and the derived 2-fold tensor product quantum codes with non-trivial 2-copy-cup gates.''
        $\xrightarrow{Enhanced}$
        \emph{Conduct a numerical search for non-abelian group algebra codes and the resulting 2-fold tensor product quantum codes with non-trivial 2-copy-cup gates.}
    \end{quote}
    \begin{quote}
        ``A further extension would be to construct cube complexes from non-abelian groups, to obtain quantum codes with non-trivial 3-copy-cup gates; this can be generalised to higher-dimensional complexes.''
        $\xrightarrow{Enhanced}$
        \emph{Generalise the construction of complexes from non-abelian groups to higher-dimensional complexes beyond cube complexes.} (Here is an example of a partial failure; In the current simple implementation, the process failed to recognize the quote discusses two different extensions).
    \end{quote}
    
\begin{figure*}
      \centering
      \includegraphics[width=1\textwidth]{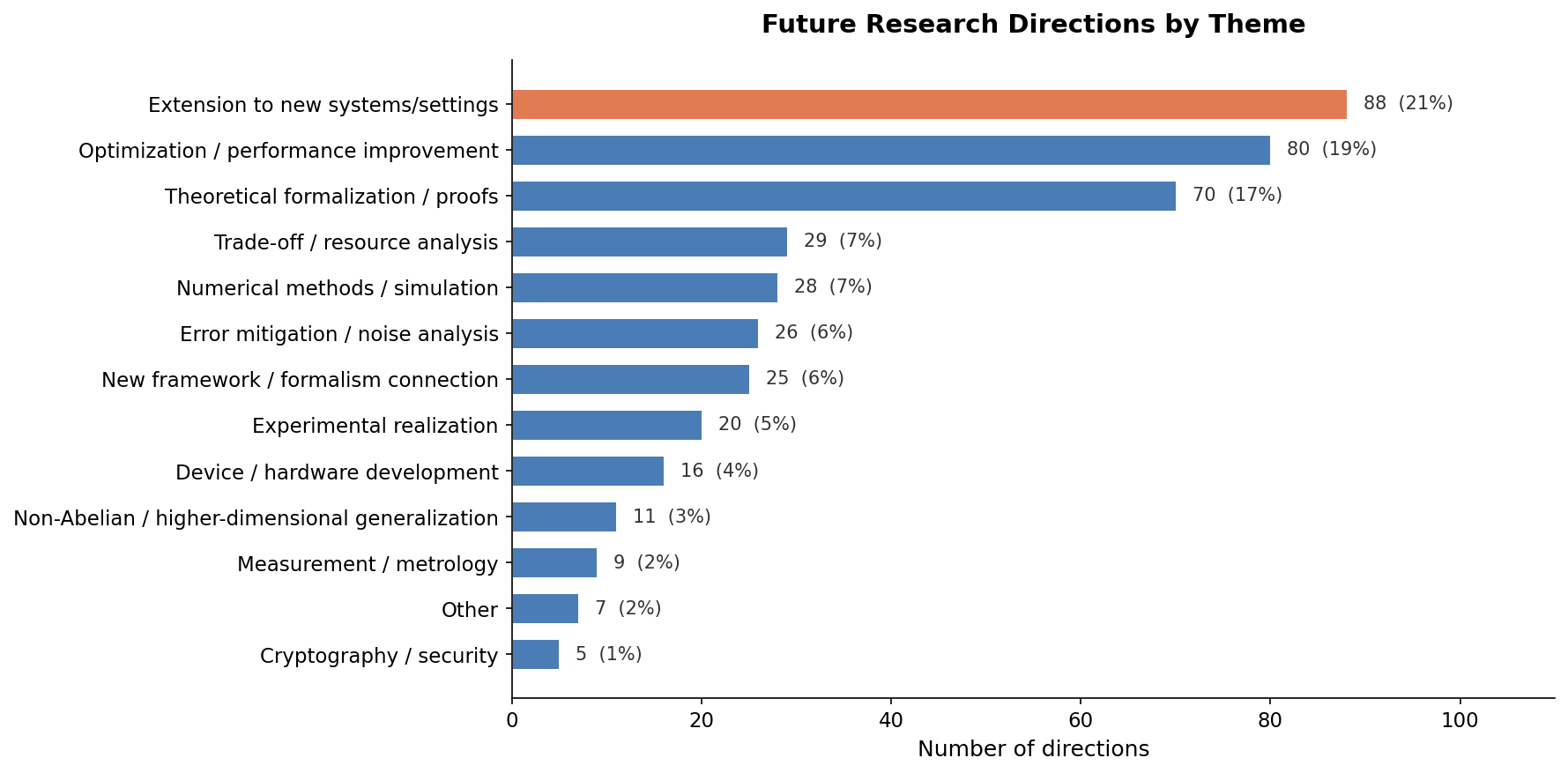}
      \caption{Major themes of \emph{future extensions} extracted from recent \texttt{quant-ph} ArXiv papers.}
      \label{fig:metacategories}
    \end{figure*}

    Next, we clustered the results and identified some major meta-categories. In Fig.\ \ref{fig:metacategories}, we show the contributions to each category (the results only show top-1 match; we did not test whether a given future direction can be fitted to multiple categories at once).

    While the pipeline still requires improvement (e.g., adding a verification loop, aligning future directions with multiple categories, filtering out papers that have already been discussed elsewhere, etc.), this simple test shows that building such a dynamic database of concepts is feasible (cf. Ref.~\cite{2405.17044}).

\section{Conclusion}

    Benchmarking agentic scientific AI systems requires measuring not only retrieval and computation, but also reasoning, tool orchestration, and critical scientific judgment. This report proposes a framework organized around four benchmark families (replication, error detection, scientific reasoning, and discovery-style tasks). It discuss also several construction strategies (planted structures, constraint modification, difficulty scaling, self-consistency, dialogue, and adversarial inputs), as well a system evaluation protocol (ablation studies, tool-use analysis, process-level evaluation, and scaling curves).
    
    Much remains open: how to evaluate creative results automatically, how to align automatic scoring methods with human judgment, how to construct difficulty ladders for open-ended scientific tasks, how to propagate errors (relevant to the error-planning strategy) through the paper in a self-consistent manner, etc.

%%%%%%%%%%%%%%%%%%%%%%%%%%%%%%%%%%%%%%%%%%%%%%%%%%%%%%%%%%%%%%%%%%%%%%%%%%%%%

\bibliographystyle{alpha}
\bibliography{bibliography}  % Produces the bibliography via BibTeX.

\end{document}